\documentclass[conference]{IEEEtran}
\IEEEoverridecommandlockouts
\usepackage{cite}
\usepackage{amsmath,amssymb,amsfonts}
\usepackage{algorithmic}
\usepackage{graphicx}
\usepackage{textcomp}
\usepackage{xcolor}
\def\BibTeX{{\rm B\kern-.05em{\sc i\kern-.025em b}\kern-.08em
    T\kern-.1667em\lower.7ex\hbox{E}\kern-.125emX}}
\begin{document}

\newcounter{example}[section]
\newenvironment{example}[1][]{\refstepcounter{example}\par\medskip
   \noindent \textbf{Example~\theexample. #1} \rmfamily}{\medskip}

\newtheorem{SampleEnv}{Sample Environment}[section]
\newcommand{\round}[1]{\ensuremath{\lfloor#1\rceil}}

\title{Hide Secret Information in Blocks: Minimum Distortion Embedding\\
}

\author{\IEEEauthorblockN{Md Amiruzzaman}
\IEEEauthorblockA{{Kent State University} \\
Kent, OH, USA \\
mamiruzz@kent.edu}
\and
\IEEEauthorblockN{Rizal Mohd Nor}
\IEEEauthorblockA{{Int'l Islamic University Malaysia}\\
Selangor, Malaysia\\
rizalmohdnor@iium.edu.my}
}

\maketitle

\begin{abstract}
In this paper, a new steganographic method is presented that provides minimum distortion in the stego image. The proposed encoding algorithm focuses on DCT rounding error and optimizes that in a way to reduce distortion in the stego image, and the proposed algorithm produces less distortion than existing methods (e.g., F5 algorithm). The proposed method is based on DCT rounding error which helps to lower distortion and higher embedding capacity.
\end{abstract}


\section{Introduction}
Often, steganography is categorized as art to hide secret messages to a medium \cite{amiruzzaman2010concurrent, amiruzzaman2009improved}, such as images, videos, etc. Steganalysis is the technique of detecting hidden secret messages from the steganographic algorithm \cite{amiruzzaman2009improved, thangadurai2014analysis}. Unfortunately, as new steganographic algorithms are developed so does the steganalysis techniques to detect the information about the hidden message. The focus of steganographic algorithms is to hide the existence of the message; on the other hand, the focus of steganalysis algorithms is to reveal the secret messages. Therefore, previous studies suggested that researchers should consider existing steganalysis techniques while they are trying to develop a steganographic algorithm \cite{amiruzzaman2010concurrent, amiruzzaman2009improved}.\\

In cryptography, the secret message is encrypted so that attackers or unwanted parties cannot read the message \cite{AmirThesis11}. However, if an attacker decrypts the secret message then the cryptographic system is broken \cite{thangadurai2014analysis}. On the contrary, in steganography, if an attacker reveals the existence of the concealed secret message then the purpose of steganography is defeated. Therefore, in steganography concealing the existence of the secret message is more important than not making the encryption more difficult to break \cite{alotaibi2019secure}. In other words, the main focus of cryptography is to protect the information from reading from unwanted parties, and steganography is to hide the existence of the information from unwanted parties.\\

Images, videos, text files, pdf files, etc. are the most common medium of steganography. However, images are the most popular medium for steganography \cite{AmirThesis11}. Because image steganography is simpler and provides comparatively higher hiding/embedding capacity \cite{bender1996techniques}. In image steganography, the Least Significant Bit (LSB) modification method is considered as a pioneer work \cite{fridrich2001detecting}. Note that, LSB modification and LSB matching have two different application areas \cite{johnson1998exploring}. LSB modification is popular for the uncompressed domain, while LSB matching is popular for the compressed domain. It is worth to mention here that the detection processes of these techniques are also different.\\
 
Several innovative steganographic algorithms are developed within the last few decades \cite{lu2019binary}. Such as, matrix embedding or F5 algorithm \cite{steganalysis2001f5}, modified matrix embedding\cite{kim2006modified}, BCH coding\cite{sachnev2009less}, and trellis-coded \cite{filler2010minimizing}. While these methods have claimed to have better resistance against statistical attacks or statistical steganalysis, many researchers have developed methods to break them.\\

Westfeld \cite{steganalysis2001f5} mentioned that statistical attack or steganalysis is the most popular attack in steganography. Hence, researchers should check steganographic algorithms against statistical attacks first \cite{amiruzzaman2010concurrent, amiruzzaman2009improved}. There are some other steganalysis techniques, such as calibrated statistics attack \cite{fridrich2003quantitative, fridrich2004feature}, which should be considered as well. However, nowadays most steganographic techniques made sure that they have resistance against statistical attacks, such as F5 algorithm \cite{steganalysis2001f5}, modified matrix embedding \cite{kim2006modified}, and secure steganographic algorithm \cite{amiruzzaman2008secure}, etc.\\

It is important to develop a steganographic algorithm by making it secure against statistical attacks; more specifically first-order statistical attack \cite{amiruzzaman2008secure}, or to keep the histogram of the JPEG image coefficients intake or less altered. In general, the JPEG image coefficient histogram is bell-shaped \cite{fridrich2002steganalysis}. Therefore, several steganographic algorithms attempted to keep the JPEG image coefficient histogram in bell-shaped after hiding the data. For example, OutGuess \cite{provos2001defending}, F5 algorithm \cite{steganalysis2001f5}, secure steganography \cite{amiruzzaman2008secure}. However, to keep the stego image (i.e., altered image) histogram may harm the quality of the stego image or may cause higher distortion \cite{amiruzzaman2008secure, steganalysis2001f5, provos2001defending}.\\

To minimize the distortion of the stego image, Kim et al. have developed the Modified Matrix Embedding (MME) algorithm \cite{kim2006modified}. In their work, the authors considered JPEG images as their secret medium. The MME algorithm helps to identify which LSB bit to modify in order to hide a secret message, this part is the same as the F5 algorithm \cite{steganalysis2001f5}. In general, modifying LSB of an image using the MME algorithm produces the least amount of distortion \cite{kim2006modified}. Yet extensive analysis could reveal the existence of a secret message because of LSB modification.\\

In a study, Pevny and Fridrich \cite{pevny2007merging}, explained JPEG image distortion because of rounding operation during the image compression. During image compression, an image goes through various operations. Such as Discrete Cosine Transform (DCT), quantization, etc. The detail of JPEG image compression is explained in \cite{AmirThesis11}. Among all the processes in JPEG compression rounding operation occurs to transform to quantized coefficients to integer numbers, which is also known as JPEG coefficients.\\   

Pevny and Fridrich \cite{pevny2007merging} described how rounding operation contributes to image distortion. The authors have explained how to reduce rounding error and ultimately distortion in the compressed JPEG image. Kim et al. \cite{kim2006modified} used the idea of reducing the rounding error in matrix embedding or F5 algorithm. The MME algorithm is well-known for its less distortion compared to the F5 Algorithm.\\

Although, the MME algorithm outperforms the F5 algorithm by minimizing the rounding error. However, because of the position of candidate coefficient is defined by the matrix embedding technique without any flexibility. So, the error minimization is yet to be the least. The proposed technique allows finding the best candidate that will allow minimizing the rounding error and overcomes the drawback of the MME algorithm.\\

The proposed technique uses a block of coefficients to hide a single bit of secret message. It modifies only one of the coefficients from the group of coefficients (i.e., whichever produces the least amount of distortion after hiding the secret message is the candidate coefficient). This method uses a block of coefficients and considers minimizing the distortion in embedding, thus the method is called Minimum Distortion Embedding (MDE).\\

The rest of this paper is organized as follows: Section \ref{lit} presents related works. Section \ref{proposed} describes the proposed method in detail, including encoding and decoding techniques. The experimental results and comparisons of obtained results are presented in \ref{results} section. Finally, section \ref{conclusion} concludes the study and provides future research directions.

\section{Existing methods}\label{lit}
\subsection{JPEG image}
During the JPEG image compression, an uncompressed image goes through JPEG encoder and decoder while transformation happens using the DCT \cite{AmirThesis11}. In the encoder, each channel of the image is divided into $8\times8$ blocks, which is also known as the JPEG block. Let, $f(i, j)$, where $i, j = 0, 1,\cdots, N-1$ of a $N\times N$ image channel block and $F(i, j)$, where $i, j = 0, 1,\cdots, N-1$ of a DCT transformation of the $N\times N$ image channel block \cite{amiruzzaman2010concurrent}.\\

Note that, the first coefficient after DCT transformation $F(0, 0)$ is known as DC coefficient and the rest of the 63 coefficients of an $8 \times 8$ block are known as AC coefficients \cite{AmirThesis11}. If the quantization matrix is denoted as $Q$, then after quantization and before rounding the coefficients can be expressed as,

\begin{equation}\label{eq:qunat}
F^{\prime}(i,j) = \frac{F(i,j)}{Q(i,j)}
\end{equation}

and after rounding the coefficients becomes integer \cite{rosenholtz1996perceptual} as described in Eq. \ref{eq:round}. 
\begin{equation}\label{eq:round}
F^{\prime\prime}(i,j) = \round{F^{\prime}(i,j)}
\end{equation}

clearly there is a difference between $F^{\prime}(i,j)$ and $F^{\prime\prime}(i,j)$ because of the rounding operation, which can be expressed as, 
\begin{equation}\label{eq:r}
r_i = F^{\prime}(i,j) - F^{\prime\prime}(i,j)
\end{equation}

\subsection{F5 algorithm}\label{sec:f5}
The F5 steganographic method was proposed by Westfeld \cite{steganalysis2001f5}. This method is considered as one of the first methods of data hiding that provides less modification during the data hiding process. The F5 algorithm is based on the matrix encoding technique, where among $u$ non-zero AC coefficients are considered to hide $v$ secret message bits by modifying only one coefficient. For example, $u = 7$ non-zero coefficients will be considered to hide $v = 3$ message bits. Since secret bits are either 0 or 1, thus $u$ and $v$ can be computed as,
\begin{equation}\label{eq:f5}
u = 2^v-1
\end{equation}
where $u$ is the number of non-zero coefficients and $v$ is the number of secret message bits.\\

It is important to note that the F5 algorithm is not breakable using the first-order statistics or histogram-based steganalysis \cite{steganalysis2001f5}. Histogram-based steganalysis compares the histogram of original image (i.e., unaltered image) and stego image (i.e., modified image) \cite{lu2019binary}. Hence, the F5 method is considered as one of the best steganographic methods \cite{AmirThesis11}. However, the F5 method has its limits as well; for example, the F5 method does not provide freedom to select a position of the candidate coefficient. In other words, the position of the candidate coefficient is not changeable.\\

Also, the F5 algorithm increases the number of zeros coefficients in the stego image and does not have the ability to minimize the distortion in the stego image \cite{AmirThesis11}. However, it important to note that the F5 algorithm is much secure than most algorithms when it hides a small amount of secret message in a JPEG image, and it is secure against the Chi-Square ($\chi^2$) analysis \cite{rezaei2019impact,AmirThesis11}.\\

\begin{example}
Let, the non-zero AC coefficients LSBs are denoted by $C = [C_1, C_2, \cdots, C_n]$, where, $n \in \mathbf{Z}$, secret message bits are denoted by $b = [b_1, b_2, \cdots, b_n]$ where, $n \in \mathbf{Z}$. Suppose there are 3 bits of secret message, and their combination can be expressed using $H$ matrix as:\\
\[H = \begin{vmatrix}0 & 0 & 0 & 1 & 1 & 1 & 1\\ 0 & 1 & 1 & 0 & 0
& 1 & 1\\1 & 0 & 1 & 0 & 1 & 0 & 1\end{vmatrix}\]

Since the coefficient matrix is one-dimensional, thus in order to multiply that with the $H$ matrix, it is important to transpose $C$. Note that only non-zero $C$ would be considered for this process. See the following Eq. \ref{eq:Coef} for more details:
\end{example}

\begin{equation}\label{eq:Coef}
H.C^T
\end{equation}

\subsubsection{Encoding of F5 algorithm}
To get the position ($p_i$) of the candidate coefficient or the coefficient that needs to change to hide secret message bits can be found using Eq. \ref{eq:bi}. This means that $b_i$ needs to be subtracted from the results of Eq. \ref{eq:Coef}. After getting the position of the candidate coefficient 1 would be subtracted if the coefficient is positive and 1 would be added if the coefficient is negative.
\begin{equation}\label{eq:bi}
p_i = b_i - H.C^T
\end{equation}

\subsubsection{Decoding of F5 algorithm}
From the stego image all non-zero coeffients would be extracted as $C^{\prime} = [C^{\prime}_1, C^{\prime}_2, \cdots, C^{\prime}_n]$, and extracted coeffients will be multiplied with $H$ matrix to get the secret message bit $b_i$. See the following Eq. \ref{eq:Coef'} for more details:

\begin{equation}\label{eq:Coef'}
b_i = H.C^{{\prime}T}
\end{equation}

\begin{example}
Suppose, there are 7 non-zero AC coefficients as [5~2~3~1~-2~-5~-1] and
corresponding LSB bits are $C$ = [5~2~3~1~-2~-5~-1], and secret message bits are $b_i$ = [1~0~1].\\

Therefore,
\[H.C^T =H \times \begin{vmatrix}5\\2\\3\\1\\-2\\-5\\-1\end{vmatrix}\]

so,\\
\[H.C^T =\begin{vmatrix}-7\\-1\\5\end{vmatrix} \mod_2 = \begin{vmatrix}1\\1\\1\end{vmatrix} \]

and the position
of the coefficient is 
\[p_i = b^T_i - H.C^T =\begin{vmatrix}1\\0\\1\end{vmatrix}- \begin{vmatrix}1\\1\\1\end{vmatrix} \]

or,\\
\[p_i =\begin{vmatrix}0\\1\\0\end{vmatrix}\]

Thus, using the $H$ matrix,  it is easy to find that second coefficient $C-2 = 2$ need to modify, so, the after modification the coefficients became $C^{\prime} = [5~1~3~1~-2~-5~-1]$.\\

To decode the message 
\[ b^T_i = H.C^{{\prime}T}\]

or, \\

\[b^T_i = \begin{vmatrix}0 & 0 & 0 & 1 & 1 & 1 & 1\\ 0 & 1 & 1 & 0 & 0
& 1 & 1\\1 & 0 & 1 & 0 & 1 & 0 & 1\end{vmatrix} \times \begin{vmatrix}5\\1\\3\\1\\-2\\-5\\-1\end{vmatrix} \]
and\\
\[b^T_i = \begin{vmatrix}-7\\-2\\5\end{vmatrix} \mod_2 = \begin{vmatrix}1\\0\\1\end{vmatrix} \]
\end{example}

\subsection{Modified Matrix Embedding (MME)}\label{sec:mme}
The MME algorithm works the exact same way as the F5 algorithm (explained in section \ref{sec:f5}). Except, after finding a candidate coefficient's position, MME alters the coefficient with the help of Eq. \ref{eq:r} and modify the coefficient using the Eq. \ref{eq:mme}.
\begin{equation}\label{eq:mme}
C^{\prime}_i = \left\{
\begin{array}{l l}
 -2, \text{ if} ~r_i \leq  0, \text{ and}~ C_i = -1\\
  C_i + 1, \text{ if} ~r_i \leq 0, \text{ and}~ C_i \neq -1\\
  2, \text{ if} ~r_i > 0, \text{ and}~ C_i = 1\\
  C_i - 1, \text{ if} ~r_i > 0, \text{ and}~ C_i = 1\\\end{array} \right.
\end{equation}

Again, it is important to note that MME can reduce the distortion by modifying the candidate coefficient in a way that provides the least distortion for that particular coefficient.

\section{Proposed MDE method}\label{proposed}
The proposed method is very simple, it uses the idea of MME as explained in section \ref{sec:mme}. However, it allows finding the candidate which provides the least amount of distortion. In other words, it has the ability to overcome the limitation of the MME algorithm. Therefore, mathematically and theoretical this method can outperform both F5 and MME algorithms.\\

In its first step, the proposed method gathers all the non-zero AC coefficients in a single array ($D_i \in \mathbf{Z}$), and divides the array into small coefficient blocks $B_i$, where $i$ = 1, 2, 3 $\cdots$, $z$, $z$ is total number of block or total number of secret message bits ($b_i$). So, The $B_z$ can be obtain by dividing the coefficient array by the number of secret message bits (see Eq. \ref{eq:block}). Let, the number of secret message bits be $\alpha$ then,
\begin{equation}\label{eq:block}
B_z = \lfloor\frac{D_n}{\alpha}\rfloor
\end{equation}

The second step is to find the coefficient that produces the least distortion in the block $B_i$

\begin{equation}\label{eq:min}
C_{min} = \min\{r_i, \{r_i, C_i \in B_i \}\}
\end{equation}
In step three, after finding the best candidate in the block the proposed algorithm modifies the coefficient following the rule explained in Eq. \ref{eq:mme}. 

\begin{example}
Suppose, a block is [5~2~3~1~-2~-5~-1], so, after adding all the coefficients the sum becomes 3, which is an odd number. So, if this block needs to hide 1 as a secret message bit, then nothing needs to be done for this case.\\

However, if this block needs to hide 0 as a secret message bit, then the sum value needs to be modified to an even number (i.e., 2 or 4). This can be done by modifying any of the coefficients. Either add 1 or subtract one would do the trick. However, because this method tries to reduce the distortion as much as possible, therefore it looks for the coefficient and $\pm1$ to it.
\end{example}

\subsection{Encoding of MDE algorithm}
The encoding scheme is very simple and easy to implement. The encoding
scheme working as follows:\\
\begin{enumerate}
\item[{(}1{)}] Make $B_i$ same as the hidden number of hidden message
bits. Make the sum of all coefficients of blocks. If the sum is odd then that sum
can represent hidden bit 1, and if the sum is even then that can
represent hidden message bit 0.
\item[{(}2{)}] Modify one of the coefficients of the block following the less distortion rule (if necessary).
\end{enumerate}

\subsection{Decoding of MDE algorithm}
The decoding scheme working as follows:\\
\begin{enumerate}
\item[{(}1{)}] If the sum of a block $B_i$ is odd then the hidden secret message bit is 1, and if the sum is even then hidden secret message bit is 0.
\end{enumerate}

\begin{example}
Suppose a block $B_i$ size is 5, and non-zero AC
coefficients are before rounding -0.6994, 0.8534,
1.7352, 1.6229, -2.6861, and after rounding the DCT coefficients became as -1, 1, 2, 2, -3.\\

So, for the given block $B_i$ the rounding errors would be as, -0.3006, 0.1466, 0.2648, 0.3771, -0.3139. Now, if modifications made by following the Eq. \ref{eq:mme}, then because of might look like as (-1-1), (1+1), (2-1), (2-1), (-3+1). Then, the error between original coefficients (before rounding) and modified coefficients (after modifying) would be -1.3006, 1.1466, -0.7352, -0.6229, and 0.6861. So, clearly the best candidate is the second to last coefficient.\\

So, the proposed method will find the second to last coefficient because of it is producing the least amount of distortion among all the coefficients, and modify it to hide the secret message bit.
\end{example}

\section{Compare the distortion between F5/MME and MDE}
Both F5 and MME use matrix embedding, which is explained in sections \ref{sec:f5} and \ref{sec:mme}. However, MME uses rounding error information to minimize the distortion using the Eq. \ref{eq:mme}. In general, there are 50{\%} chances that the MME method reduced the distortion than the F5 algorithm. Perhaps, the F5 algorithm modified a coefficient in a way that produces the same distortion as MME. Also, this can be proved by the logic of uniform distribution \cite{kuipers2012uniform}. As the secret message bits are either 0 or 1, thus in most cases the coefficients did not need to be modified as they may already be in a form that can represent the secret message bit \cite{yang2018roi}.\\ 

In addition, both F5 and MME get the position of the candidate coefficient to modify. There may be another coefficient that may produce less distortion than the candidate coefficient, however, F5 and MME method do not get to pick that coefficient to modify and hide data. In contrary, the purposed MDE method find which coefficient produces the least distortion $C_c = \min\{C_i \in B_i|\min(r_i)\}$, where $C_c$ is the candidate coefficient that produces the least amount of distortion after hiding the secret message bit.\\

Therefore, based on the above discussion it is clear that MDE method outperform both F5 and MME methods, and produces less distortion than those methods. In sum, MDE produces less error so probability of detecting the existence secret message in MDE is less than F5. The following section would show some experimental comparisons between MDE and F5 method as it it the base method.\\

\section{Experimental Results and Comparisons}\label{results}
\subsection{Analysis using error probability}
An image database was used to test both the F5 method and the proposed MDE method. In the database, there are over 1,173 images. A Support Vector Machine was trained using the features of original images from the database. Then tested by modifying those images using F5 and MDE methods. Results of F5 and MDE were checked separately to compare the error probability. If SVM produced more error probability for one method than the other that showed which method is stronger or have better strength against steganalysis \cite{fridrich2007statistically}.\\

The \textit{error probability} can be defined using following equation:
\begin{equation}\label{eq:ep}
P = \frac{P_{FA}+P_{MD}}{2}
\end{equation}
where, $P$ is the error probability, $P_{FA}$ is the false positive and $P_{MD}$ missed detection. Note that, if a method reaches 50{\%} error probability, then that would mean that SVM is unable to determine whether there is any secret message hidden in it. Therefore, 50{\%} error probability is desired.\\ 
The proposed method and F5 method compared with two different JPEG image quality factors (QF), such as QF = 50 and QF = 75. Information loss happens during the image compression process, and this loss of information is measured by a quality factor or QF \cite{butora2019effect}.  A higher-quality factor means less information loss. For example, QF = 50 means more information loss than QF = 75. In the case of QF = 50, results suggested that the F5 algorithm produced smaller steganalysis error probability than the proposed method MDE algorithm. For example, with 5{\%} data hiding capacity F5 algorithm produced 23.085 steganalysis error probability and the proposed MDE method produced 44.8041. Again, a higher value of error probably means less detectable.\\ 

Likewise, with 10{\%} data hiding capacity and with QF = 50, F5 algorithm produced 4.5997 steganalysis error probability, while the proposed MDE method produced 33.0494 steganalysis error probability. Then again, with 15{\%} data hiding capacity and with QF = 50, F5 algorithm produced 2.0443 steganalysis error probability, and the proposed method produced 18.9949 steganalysis error probability. In addition, with 20{\%} data hiding capacity and with QF = 50,
F5 algorithm produced 0.5111 steganalysis error probability, and the
proposed MDE method produced 4.4293 steganalysis error probability (see Table \ref{table}).\\ 

Similarly, with 5{\%} data hiding capacity and QF = 75, F5 algorithm produced 18.3986 steganalysis error probability, and the proposed MDE method produced 44.9744 steganalysis error probability. With 10{\%} data hiding capacity and with QF = 75, F5 algorithm produced 2.1995 steganalysis error probability, and the proposed MDE method produced 33.3049 steganalysis error probability. With 15{\%} data hiding capacity and with QF = 75 F5 algorithm produced 0.6814 steganalysis error probability, and the proposed MDE method produced 17.4617 steganalysis error probability. Again, with 20{\%} data hiding capacity and with QF = 75 F5 algorithm produced 0.2555 steganalysis error probability, and the proposed method produced 3.8330 steganalysis error probability
(see Table \ref{table}).\\
\begin{table}
\caption{Error probability comparison} \label{table}
 \centering
\begin{tabular}{|c|c|c|c|c|c|}
\hline
\multicolumn{5}{c}{} Steganalysis by Error Probability (EP) \\

\hline \hline
 &  & 5\% & 10\% & 15\% & 20\%\\
\hline
QF = 50 & F5 & 23.0835 & 4.5997 & 2.0443 & 0.5111\\
\cline{2-6}
     & MDE & 44.8041 & 33.0494 & 18.9949 & 4.4293\\
     \hline
     QF = 75 & F5 & 18.3986 & 2.1295 & 0.6814 & 0.2555\\
\cline{2-6}
     & MDE & 44.9744 & 33.3049 & 17.4617 & 3.8330\\

\hline
\end{tabular}
\end{table}

\subsection{Analysis using embedding rate}
The hiding rate or \textit{embedding rate} of a image can be computed as \cite{AmirThesis11, taburet2019natural},

\begin{equation}\label{eq:hidingrate}
\mbox{Embeding rate} = \frac{\mbox{Number of secret bits}}{\mbox{Capacity of encoding}}\times 100
\end{equation}
Both MDE and F5 methods were tested with Support Vector Machine to detect steganalysis probability, the following comparison are prepared after getting the steganalysis detection result.
During the performance testing, the error probability and embedding rate were considered with both QF = 50, and QF = 75. With both QF (i.e., 50 and 75), the proposed method (i.e., MDE) achieved better performance than the F5 method.
A distortion function was $D(.)$ was defined by the guidelines provided by Filler and Fridrich \cite{filler2010minimizing} as below:

\begin{equation}\label{eq:distortionfunction}
D(C^{\prime}, C) = \sum_{i=1}^{n}\rho_i(c^{\prime}_i, c_i)
\end{equation}

where $\rho_i:C \rightarrow \mathbb{R} \cup \{\infty\}$ is the cost functions satisfying $\rho_i(c^{\prime}_i, c_i) = \infty$ whenever $c_i \neq c^{\prime}_i$.\\

A steganography method will have better resistance if it produces less mean distortion \cite{fridrich2013multivariate, yao2015defining}. During the performance testing, the mean distortion and embedding rate was considered with QF = 50, and QF = 75. With both QF, the proposed MDE method has achieved better performance than F5 method.\\

\subsection{Visual analysis}
\begin{figure}
\begin{center}
\includegraphics[width=7cm]{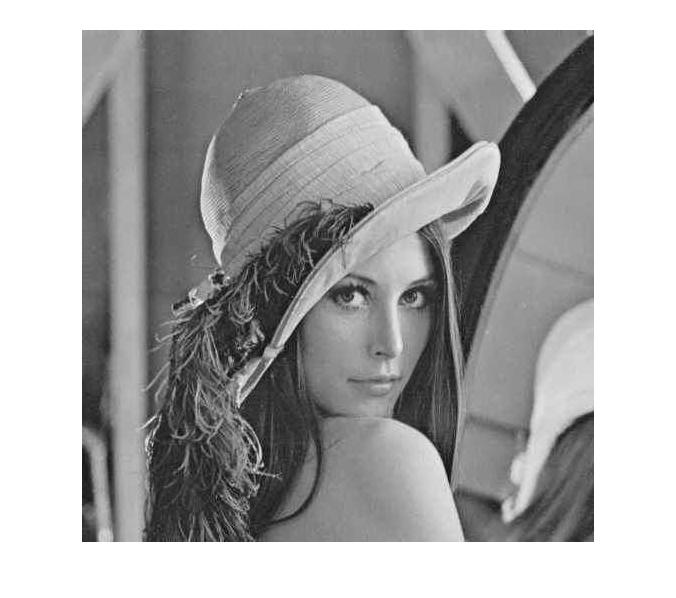}
\caption{Lena image after embedding by
proposed method.}\label{lena}
\end{center}
\end{figure}

\begin{figure}
\begin{center}
\includegraphics[width=7cm]{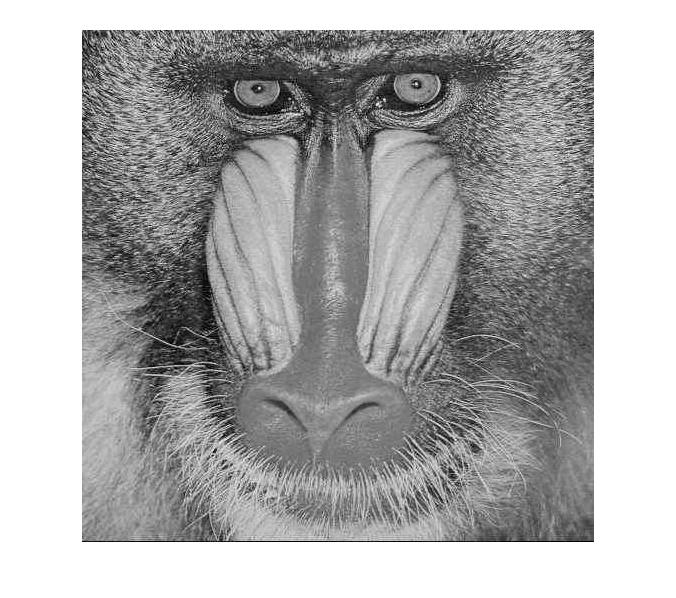}
\caption{Baboon image after
embedding by proposed method.}\label{baboon.jpg}
\end{center}
\end{figure}

\textit{Visual analysis} is another important steganalysis technique that helps to understand whether an image was artificially altered \cite{wang2004cyber, cheddad2008enhancing}. As the proposed method produces less distortion than the F5 method. Therefore, the output image or stego image produced by proposed MDE method does not show any visual issues \cite{AmirThesis11}, and it will not reveal the existence of secret message in visual inspection (see Fig. \ref{lena} and Fig \ref{baboon.jpg}).\\

\section{Conclusions}\label{conclusion}

The results of this study suggest that the proposed MDE method has better resistance ability in terms of steganalysis than the F5 method. In the case of steganography, attacks are more important than capacity, while this method has better hiding capacity also \cite{AmirThesis11}. The main advantage of this proposed method is the freedom of modifying any coefficients. Resulting in a better quality of stego image and higher resistance against attacks.\\

As for the future study, other steganographic algorithms should be considered for the comparison. In addition, more feature-based steganalysis should be considered as more features may increase the chance of detection \cite{fridrich2011steganalysis}. Moreover, a study could find a way to increase the hiding capacity and yet maintain less distortion ability.

\bibliographystyle{IEEEtran}
\bibliography{IEEEabrv, bibfile}

\end{document}